# New high-tech flexible networks for the monitoring of deep-sea ecosystems

**Running Head: High-tech deep-sea monitoring networks**

**Aguzzi Jacopo[1,*], Chatzievangelou Damianos[2], Marini Simone[3], Fanelli Emanuela[4], Danovaro Roberto[4,5], Flögel Sascha[6], Lebris Nadine[7], Juanes Francis[8], De Leo Fabio [8,9], Del Rio Joaquin[10], Thomsen Laurenz[2], Costa Corrado[11], Riccobene Giorgio[12], Tamburini Cristian[13], Lefevre Dominique[13], Gojak Carl [14], Poulain Pierre-Marie[15], Favali Paolo[16,17], Griffa Annalisa[3], Purser Autun[18], Cline Danelle[19], Edgington Duane[19], Navarro Joan[1], Stefanni Sergio[5], D'Hondt Steve[20], Priede G. Imants [21,22], Rountree Rodney[8,23], Company B. Joan[1]**

[1,*] Instituto de Ciencias del Mar (ICM-CSIC). Paseo Marítimo de la Barceloneta, 37-49. 08012. Barcelona, Spain. Tel: +34 93 230 95 00, Fax: +34 93 230 95 55, mail: jaguzzi@icm.csic.es

[2] Jacobs University, Bremen, Germany

[3] National Research Council of Italy (CNR), Institute of Marine Sciences, La Spezia, Italy

[4] Department of Life and Environmental Sciences, Polytechnic University of Marche, Ancona, Italy

[5] Stazione Zoologica Anton Dohrn (SZN), Naples, Italy

[6] GEOMAR, Kiel, Germany

[7] Sorbonne University, CNRS LECOB, Oceanological Observatory, Banyuls-sur-mer, France

[8] Department of Biology, University of Victoria, Victoria, Canada

[9] Ocean Networks Canada (ONC), University of Victoria, Victoria, Canada

[10] OBSEA, SARTI, Universitat Politècnica de Catalunya (UPC), Barcelona, Spain

[11] Consiglio per la ricerca in agricoltura e l'analisi dell'economia agraria (CREA-IT), Monterotondo, Italy

[12] Istituto Nazionale di Fisica Nucleare (INFN), Laboratori Nazionali del Sud, Catania, Italy

[13] Institut Méditerranéen d'Océanologi.e. (MIO), Marseille, France

[14] DT INSU, La Seyne-sur-Mer, France

[15] Istituto Nazionale di Oceanografia e Geofisica Sperimentale (OGS), Trieste, Italy

[16] Istituto Nazionale di Geofisica e Vulcanologia (INGV), Rome, Italy

[17] European Multidisciplinary Seafloor and water-column Observatory European Research Infrastructure Consortium (EMSO ERIC), Rome, Italy

[18] Alfred Wegener Institute (AWI). Bremerhaven, Germany

[19] Monterey Bay Aquarium Research Institute (MBARI), Moss Landing, CA, USA

[20] Graduate School of Oceanography, University of Rhode Island, RI, USA

[21] University of Aberdeen, Aberdeen, UK.

[22] Hellenic Centre for Marine Research, Heraklion Crete, Greece

[23] The Fish Listener, 23 Joshua Lane, Waquoit, Massachusetts, USA

## ABSTRACT

Increasing interest in the acquisition of biotic and abiotic resources from within the deep sea (e.g. fisheries, oil-gas extraction, and mining) urgently imposes the development of novel monitoring technologies, beyond the traditional vessel-assisted, time-consuming, high-cost sampling surveys. The implementation of permanent networks of seabed and water-column cabled (fixed) and docked mobile platforms is presently enforced, to cooperatively measure biological features and environmental (physico-chemical) parameters. Video and acoustic (i.e. optoacoustic) imaging are becoming central approaches for studying benthic fauna (e.g. quantifying species presence, behaviour, and trophic interactions) in a remote, continuous, and prolonged fashion. Imaging is also being complemented by *in situ* environmental-DNA sequencing technologies, allowing the traceability of a wide range of organisms (including prokaryotes) beyond the reach of optoacoustic tools. Here, we describe the different fixed and mobile platforms of those benthic and pelagic monitoring networks, proposing at the same time an innovative roadmap for the automated computing of hierarchical ecological information of deep-sea ecosystems (i.e. from single species' abundance and life traits, to community composition, and overall biodiversity).

## 1. Introduction

Throughout the Anthropocene Era (Crutzen & Steffen, 2003) the human footprint on the ecosystems of the global ocean has been increasing continuously (Halpern et al., 2015). As this footprint is rapidly expanding toward great depths, the need for a global observing effort in the deep ocean is crucial (Levin & Le Bris, 2015). The accurate monitoring of our incursive impacts on marine ecosystems, however, requires the development of novel and effective technological solutions.

The deep-sea seafloor and overlying waters (below 200 m depth) form the largest biome on Earth, although it remains poorly explored (Ramirez-Llodra et al., 2010; Mora et al., 2011; Sutton et al., 2017). The monitoring of key ecosystem features and functions has proven difficult, owing to the extreme environmental conditions associated with these depths (e.g. high pressures, low temperatures, corrosiveness and remoteness), coupled with limited sampling capabilities offered by low numbers of adequately equipped research vessels (Woodall et al., 2018).

In order to sustain correct management and protection actions a spatiotemporally extended monitoring regime must be implemented to gather data on species and their communities across the vast extent of the great global ocean basins (Danovaro et al., 2017). Clear examples of shortfalls in current data include the lack of knowledge on biomass, abundance, reproductive cycles, population dynamics (i.e. growth and mortality), migrations and geographic ranges (Danovaro et al., 2014). Furthermore, community biodiversity, food web structures and the influence of organic matter transfer within ecosystem compartments and across boundaries are also poorly studied in relation to the neighboring shallower and coastal ecosystems (Snelgrove et al., 2018). All these aspects have repercussions on penetration and propagation of human footprint into marine ecosystems (e.g. pollutants and microplastics; Zhao et al., 2018).

To fill these knowledge gaps, the efficient integration of ongoing technological developments into a strategic framework for deep-sea monitoring is critical (e.g. Aguzzi et al., 2012). Such development should be capable of producing tools for the spatiotemporal location and quantification of deep-sea organisms across a wide range of body sizes, as well as their activity and response to changing environmental conditions and anthropogenic stressors.

### *1.1. Objectives*

In this study, we review the status and development of high-tech, interactive networks of fixed and mobile platforms, currently used for spatiotemporally flexible and appropriate monitoring of deep-sea ecosystems. We propose an innovative roadmap for the hierarchical extraction of ecosystem indicators related to assemblage structure, biodiversity and ecosystem functioning, as obtained from biological variables encompassing species abundances, demographic descriptors, and behavior. We center our analysis on ecosystem indicators extracted from video and acoustic imaging of marine megafauna (i.e. organisms of size from centimeters and above), representing the apical ecological complexity component, that is fundamental in conditioning ecosystem functioning, services, and health (Schoening et al., 2012).

## 2. Growing high-tech cabled observatory networks

The ongoing technological development in seafloor cabled observatories is motivated by the growing awareness about the strategic value of acquiring multidisciplinary biological and environmental data in a concomitant fashion, in order to derive putative cause-effect relationships as drivers of ecosystem changes (Favali et al., 2010; 2015; Lelievre et al., 2017). The successful integration of such platforms equipped with camera systems, multi-parametric biogeochemical, oceanographic, and biological sensors with seafloor power and communication cables now allows the remote, continuous, high-frequency (> 1Hz as real-time), long-term (up to decades) monitoring of the deep-sea biome (Ruhl et al., 2011). In this highly integrated monitoring approach, megafauna identification, tracking and counting through optoacoustic and new molecular sensors should be a key focus, in relation to productivity and services (e.g. fishery; Aguzzi et al., 2015).

Throughout the last two decades, cabled observatories have provided relevant data, helping to fill the gaps in knowledge on species presence, behavior, and associated changes in biodiversity and ecosystem function (**Table 1**). Unfortunately, cabled systems are fixed and have limited spatial coverage when the deep continental margins and ocean basins are considered as a whole (Aguzzi et al., 2015; Danovaro et al., 2017). An attempt to overcome such a limitation has occurred in some cases through the installation of a local network of seabed platforms. Good examples are the Ocean Network Canada (ONC), Deep-ocean Environmental Long-term Observatory System (DELOS) and Lofoten Verlag observatory (LoVe), respectively in Juan de Fuca plate (NW Pacific), off Angola (SE Atlantic) and in Norway (Barnes et al., 2007; Vardaro et al., 2013; Bagley et al., 2015; Osterloff et al., 2016).

The deployment of observatory modules in clusters with separation distances on the order of hundreds of meters or a few kilometres is presently envisaged to maximize the ability to quantify species distributions and habitat associations over multiple scales (see **Table 1**). Deploying multiple ecosystem observatory clusters along environmental or habitat gradients would be effective in elevating the system from examination of local habitats to ecosystem level observation. Each node can acquire imaging and acoustic as well as multiparametric environmental data in a temporally coordinated fashion. Accordingly, temporal changes in species presence and abundance in an area of the deep sea can be tracked through neighboring environmental niches (Doya et al., 2017; Thomsen et al., 2017*).*

### *2.1. Permanent mobile platforms increase spatial monitoring capability*

Nevertheless, networks of cabled observatories are not enough to ensure efficient monitoring across highly variable benthic seascapes (Aguzzi et al., 2015). Presently, there is a drive to integrate mobile platforms through docking stations into existing cabled observatory infrastructures, to provide extended coverage at local, regional and basin-wide spatial scales, both on the seafloor and within the water column (**Figure 1**). Benthic mobile platforms are represented by crawlers: a new class of Internet Operated Vehicles (IOVs), tethered to cabled observatories (Purser et al., 2013). These tracked vehicles are capable of real-time navigation control and data collection *via* simple web browser interfaces operable from anywhere. At the same time, a new class of rovers, non-tethered benthic mobile crawlers are entering into active research, capable of automatically returning to the docking station for charging, data transfer or recovery (Flögel, 2015). To complement the seafloor monitoring capacities of crawler systems, pelagic monitoring is presently achieved using tethered Remotely Operated Vehicles (ROVs) and free swimming Autonomous Underwater Vehicles (AUVs) some of which may, also dock with cabled stations for energy recharge and data transmission (Bellingham, 2016). These also allow monitoring of the water column at a high-frequency over extended periods and across depth strata (e.g. Ludvigsen & Sorensen, 2016; Masmitjia et al., 2018).

Each of these mobile platforms provides a unique contribution to the ecosystem observatory, as well as some task redundancy. The AUV equipped with imaging or acoustic devices is ideally suited for habitat and biota distribution mapping (Morris et al., 2014, Williams et al., 2016) and can be used to conduct transects around the observatory and between observatories. The AUV provides the highest mobility and flexibility in sampling design for mapping with impacts on the benthic habitat by maritime activities (e.g. noise, substrate disturbance at different scales and artificial light pollution effects). Although the ROV design implies a tether, such a platform has also a high mobility and it can be used similarly to AUVs (Robinson et al., 2017), with the advantage of having two way real-time data transmission and manipulator arms to be used for management and maintenance tasks within the monitoring infrastructure (e.g. manipulative experiments or for placing autonomous recorders such as stand-alone autonomous cameras). In addition, ROVs are the best option for collecting video data on the development of the fouling community on the observatory infrastructure and fauna association with the structure. The major drawback of ROVs is that they must operate with thrusters, creating high levels of noise and their limited ability to conduct sampling and observations at specific locations for extended periods of time (Rountree & Juanes, 2010). Crawlers, on the other hand, can be used to conduct census observations at specific locations (in constant transect or stepping-stone fashion) for extended time periods (minutes to hours). Crawlers can also share some infrastructure servicing tasks with the ROVs and carry larger payloads. Drawbacks to crawlers include noise production, but more importantly physical disturbance of the benthic habitat and associated fauna along the movement tracks.

### *2.2. Benthic networks growing in the pelagic realm*

The need to monitor energy fluxes between pelagic and benthic ecosystem compartments (i.e. benthopelagic coupling) and their spatiotemporal changes (e.g. Griffiths et al., 2017), requires the development of three-dimensional monitoring networks of platforms, with cabled nodes and mobile platforms operating in tandem (**Figure 2**). This ecologically integrated monitoring is presently being facilitated by incorporating to the benthic data collection, secondary data streams supplied by water column fixed (i.e. moored) and superficial buoys, as well as satellites (Thomsen et al., 2017). Satellites are optimal tools for gathering large-scale physicochemical data from superficial (i.e. epipelagic) ecosystems, quantifying relevant biological variables from ocean color (e.g. chlorophyll content, particulate matter, etc.). Unfortunately, satellite sensors cannot penetrate much beyond the surface of global water mass, and therefore pelagic buoys are more appropriate for the monitoring of sub-surface oceanic strata.

In this scenario, benthopelagic monitoring capabilities are also being potentiated *via* data collection from the routine operations of large astrophysical experimental infrastructures, such as underwater neutrino telescopes (see **Figure 1C**). These telescopes consist of arrays of vertically moored (up to 700 m), flexible, strings or towers of photon detectors (Photo-Multiplier Tubes; PMTs) for neutrino particle quantification, placed at different altitudes above the seabed and connected to shore *via* power and fiber-optic data cables (Adrián-Martínez et al., 2016b).

Although the primary use of these platforms is within the high-energy astrophysics domain (Adrían-Martínez et al., 2016a), their infrastructure provides a network of subsea connection points and sensors usable for marine ecological monitoring. Hydrophones for passive acoustic listening are connected to the system to monitor position of the towers in relation to currents and to simultaneously triangulate PMT location with the aid of acoustic beacons, so that the trajectories of detected neutrinos can be properly computed. As a by-product, this real-time acoustic monitoring produces useful oceanographic flow condition data and information on anthropogenic marine noise, as well as cetacean movement, population structure, and communication (Nosengo et al., 2009; Sciacca et al., 2015; Viola et al., 2017). The PMT detectors themselves also provide unique high-frequency and continuous data on bioluminescence, as swimming animals luminesce when hitting the infrastructures (Aguzzi et al., 2017). At time of writing, real-time and continuous data acquisition from these telescope infrastructures as a whole is providing important information on seasonal changes in gravity carbon fluxes and controlling oceanographic processes (e.g. dense shelf water cascading and effects on deep-sea bacterial productivity; Tamburini et al., 2013; Adrián-Martínez et al., 2016b; Durrieu de Madron et al., 2017).

## 3. A roadmap for the monitoring of ecosystem indicators

The development of efficient deep-sea ecosystem monitoring is currently based on the successful extraction and quantification of key ecosystem characteristics (e.g. biogeochemistry, animal presence, abundance and behavior, local and regional biodiversity, ecosystem functioning; see **Table 1**). This monitoring development is being based on the combined use of optoacoustic and molecular biological sensors which are being implemented in the framework of cabled observatories. The capability to acquire temporally-related time series of multiparametric habitat and biological data, allows researchers to envision aspects such as, benthic primary production *via* chemosynthesis, deep-sea species ecological niches and food web structure (Aguzzi et al., 2011; 2012; 2018). These datasets can be used to feed new numerical-based ecology approaches centered on multivariate statistics, time series analysis and ecosystem modeling (e.g. Borcard & Legendre, 2002, Matabos et al., 2014; Puillat et al., 2014; Thomsen et al., 2017), in order to estimate the level of significance for putative cause - effects relationships (i.e. environmental control *versus* species and communities response) and provide an immediate vision of complex ecological processes at a local scale (e.g. species tolerance to the variation of key habitat drivers). This approach allows a transition from a still too descriptive deep-water and deep-sea ecology into a more quantitative one, as occurs in more directly accessible coastal areas and land.

To optimize the outcome quality from a highly-integrated deep-sea monitoring strategy of this type, protocols for data collection and analysis should be implemented to efficiently characterize local biodiversity along with those processes that sustain it and determine the overall ecosystem functioning and health status (Allen et al., 2008; Danovaro et al., 2016). From an operational point of view, a bottom-up scheme of monitoring should be conceived with cabled observatories and docked mobile platforms producing video and acoustic imaging information on fauna within a wide range of sizes (e.g. from macro-zooplankton to megafauna classification and counting, morphometric description, and quantification of intra- and inter-specific interactions). Then, acquired baseline biological data can be directly related to multiparametric environmental information obtained *via* the concomitant collection of geochemical and oceanographic data (Aguzzi et al., 2012; 2015; Ferrari et al., 2016).

### *3.1. The central role of optoacoustic technologies for monitoring*

High-definition still and video image data (e.g. 2D, 3D, hyperspectral) and active acoustic imaging (i.e. multi-beam cameras; Juanes, 2018) to date represent key approaches for the optoacoustic monitoring of remote deep-sea ecosystems (Danovaro et al., 2017). Outputs of optoacoustic monitoring provide relevant data for management in key human activities, as for example fisheries or jellyfish blooms (Samhouri et al., 2014, Bicknell et al., 2016, Corgnati et al., 2016; Marini et al., 2018). Moreover, species distribution and habitat use can be studied over extended spatial scales by mosaicking high-resolution imagery, captured by mobile platforms operating in the regions surrounding the cabled infrastructure stations (Purser et al., 2013) or by integrating laser-scanning systems into the mobile platforms, to create high-resolution 3D full-color surface models (GE Reports Canada, 2017). Further development of similar methods that expand the spatial coverage of (stereo)-imaging data, can help with the quantification of other biological components and fauna sizes of high ecological relevance which are more difficult to quantify remotely over extensive areas of the deep sea. Fixed cameras (Aguzzi et al., 2011) and mobile platforms (Valentine et al., 2016) can be used to assess epibenthic bacterial mat coverage in combination with customized molecular and chemical microsensors, providing *in situ* analysis of microbial communities (see Section 3.3 below), a proxy for chemosynthetic production at reducing sites (e.g. cold-seeps, hydrothermal vents; Purser et al., 2013; Russ et al., 2013).

Video imaging at depth requires continuous illumination which carries a poorly understood potential for harmful effects on deep-sea fauna (Herring et al., 1999; Irwin, 2018). However, digital still time-lapse cameras may collect *in situ* images with triggered flash illumination, limiting the exposure to light of these perpetually dark deep-sea ecosystems. At the same time, red or infra-red lighting, at wavelengths not detectable by deep-sea animals, has been used with some success (Widder et al., 2005) but those wavelengths are rapidly attenuated in water and the resulting monochrome images contain much less information than equivalent color images (Priede, 2017).

Classic high-definition video monitoring approaches are being integrated with novel acoustic imaging systems (Rountree, 2008; Juanes, 2018) with an increasing level of complementarity in deep-water areas (**Figure 3**). Acoustic cameras, such as high-frequency multi-beam imaging Dual-frequency Identification Sonar (DIDSON) and Adaptive Resolution Imaging Sonar (ARIS) can visualize fish and invertebrate shapes, and track the movement of individuals at distances greater than those which may be achieved by visual systems equipped with artificial lighting solutions (Martignac et al., 2015).

A limitation of acoustic camera use for monitoring fauna however is related to animal identification, which with acoustic systems must be solely based on morphology, since no colorimetric and limited texture information is captured by acoustic camera devices. Spatial resolution of acoustic cameras is also insufficient to resolve important details for species identification. However, acoustic cameras can effectively 'see in the dark', thus avoiding photic contamination, allowing investigation of how artificial lights may influence animal behavior in the deep sea. In order to verify identifications, acoustic cameras must be deployed simultaneously with new prototype low-light high resolution optical imaging equipment (e.g. Barbier et al., 2012).

The space sampled around an observatory can be also increased by mounting an imaging sonar on a rotating head (see **Figure 3B**). At present such devices are installed on the ONC cabled observatory in Barkley canyon. These sonars allow internet connected operators to qualitatively discern the presence and abundance of benthic fauna and any associated bioturbation over surfaces larger than in any single fixed image (Robert & Juniper, 2012). Similar rotating side or upward facing sonar packages are undergoing initial deployments on other cabled infrastructures, capable of being used to identify animals, when they are not too densely grouped, at distances of up to ~1 km (Godø et al., 2014).

Currently, automation in image processing for animal tracking, classification, counting, the extraction of morphological features (e.g. size, shape, color patterns), and characterization of behavioral aspects (e.g. crawling, walking, swimming, burying, and territoriality; *sensu* Aguzzi & Company, 2010), is becoming a relevant tool in biological data provision from cabled observatories and their associated mobile platforms. More automated routines are urgently required as the volume of image data collected by these systems increases in line with technological developments. Such routines will enable researchers to overcome the human analysis-dependent bottleneck of manual processing (Aguzzi et al., 2012), whilst also reducing observer bias (Schoening et al., 2012). By developing Artificial Intelligence (AI; in the form of learning algorithms) in computer vision, cameras may be transformed into the equivalent of a calibrated sensor, automatically providing time series quantitative data on key fauna, to augment the qualitative data represented by the images themselves (Corgnati et al., 2016, Marini et al., 2018). Despite the difficulties inherent in converting the expert knowledge into useful algorithms, calibration and tuning *via* sufficiently extensive feedback can result in operational performances comparable to those of expert researchers (MacLeod et al., 2010).

### *3.2. Passive Acoustic Monitoring (PAM) to support image-based monitoring*

PAM monitoring of fish and invertebrate sounds increase species monitoring capability well beyond the reach of optoacoustic technologies. Such a technological application has become an important tool in fisheries and conservation research (Rountree et al., 2006; Luczkovich et al., 2008). The use of PAM assets provides a long-range monitoring capability in remote locations where traditional sampling methods are difficult or impossible to implement (ACT, 2007), as for example in the case of sponge reefs (Archer et al., 2018) or seamounts (Riera et al., 2016). Furthermore, combining acoustic localization with video and other forms of observation can be used to identify sound producing species as well as document their soniferous behavior (Mouy et al., 2018). This approach is finding an increasing use in the collection of long-term data for integrated biodiversity assessment (Pieretti et al., 2017).

Unfortunately, the application of PAM is limited by the paucity of archived data on fish sounds (Rountree et al., 2002; Rountree et al., 2018a). For example, of the approximately 400 fish species in British Columbia waters, only 22 have been reported to “vocalize” in large part because sound production has been investigated in so few species (Wall et al., 2014). This is especially true in the deep sea, where fish sounds have rarely been studied despite the fact that many species possess sonic muscles presumably used in vocalization (Rountree et al., 2012; Parmentier et al., 2018).

Although many fishes and invertebrates do not produce purposeful sounds, it is important to understand that incidental sound production may occur upon physiological and behavioral activity (e.g. specific swimming and feeding mode sounds). Those acoustic marks can be used to assess the presence of individuals for a certain species and are therefore being incorporated into PAM monitoring procedures (Rountree et al., 2006; Rountree et al., 2018b).

The aforementioned PAM applications, combined with other observation technologies (e.g. video, acoustic imaging and sonar) improving the documentation of organism sound production and associated behavior, will add further ecological value to the integrated monitoring framework of ocean observatories (Rountree, 2008).

### *3.3. Molecular sensing as benchmark for species traceability*

Molecular tools have diverse applications in marine ecological studies and biological monitoring. Substantial contributions have been provided by several

DNA barcode initiatives generating and implementing databases, along with the development of metabarcoding protocols to recover community diversities from unsorted samples (Stefanni et al., 2018). The latest revolution in bio-monitoring is linked to the collection and analysis of genetic material obtained directly from environmental samples, namely environmental DNA (eDNA). This protocol enables tracing of the presence of species from skin cells, fish scales, gametes and food left overs, without the need to isolate any target organisms (Taberlet et al., 2012). Direct sequencing of eDNA has been shown to provide several advantages over traditional techniques, improving the capacity to unravel the “hidden” biodiversity (e.g. detect rare, cryptic, elusive and non-indigenous species in the early stages of invasion) and enabling global census of species in near real-time (Stat et al., 2017).

However, eDNA tracing presents some limitations such for example, the detection of false positives (when target species is absent but its DNA is recovered), and false negatives (i.e. species undetected where they are present) which have to be carefully evaluated and avoided (Taberlet et al., 2012; Cristescu & Hebert, 2018). Major difficulties encountered in deep-sea ecosystems for studies involving molecular analysis of diversity are: the general lack of taxonomic knowledge as well as the absence of appropriate databases of species-specific marker sequences (Carugati et al., 2015; Dell’Anno et al., 2015; Sinniger et al., 2016). When these molecular markers are identified (Barnes & Turner, 2016), *in situ* hybridization techniques may be used with great success when targeting expected taxa within monitoring programmes (Scholin, 2010; Berry et al., 2019).

Recent technical improvements concern the development of “Eco-genomic” sensors capable of autonomously collect biological samples and perform molecular analyses (Ottesen, 2016). These sensors allow the characterization of marine community composition as a whole, regardless of the faunal size classes involved (McQuillan & Robidart, 2017). One example is the Environmental Sample Processor (ESP Scholin et al., 2009), designed to autonomously collect discrete water samples, concentrate microorganisms, and automate the application of molecular probe technologies.

In parallel, recent advances in high-throughput sequencing technologies are allowing the processing of huge amounts of genomic data using small portable devices (i.e. miniaturized sequencers such as produced by Oxford Nanopore Technologies having the size of a USB stick). This kind of devices, together with advances in bioinformatics could represent the most important revolutionary breakthrough technology in ecological networks monitoring. Challenges related to the taxonomic assignments of genomic sequences and their interpretation (incompleteness of databases) may be solved applying machine learning algorithms (Cordier et al., 2017). Such approaches can maximize ecologically meaningful insights and provide a list of highly informative sequences ecosystem indicators that could provide the basis for hybridization chips (i.e. micro-arrays) for denser, mobile and cheaper *in situ* devices that can be scaled up appropriate spatiotemporal resolutions (Cordier et al., 2018).

AI approaches are gaining relevance in the metabarcoding analysis and provide a fast and cost-effective way for assessing the quality status of ecosystems (Cordier et al., 2018; 2019). Recent examples in -omics analysis were based on Random Forest (Breiman, 2001; Fernández-Delgado et al., 2018) and Self Organizing Maps (Kamimura, 2019). These were used for identifying biotic indices for the foraminiferal metabarcoding. Similarly, Gerhard & Gunsch (2019) used a random forest based approach for selecting the relevant biomarkers for classification of ocean, harbor and ballast water samples. LeCun, (2015) used a deep recurrent neural network (approach for a base calling application on portable sequencing machines (Merelli at al., 2018), where meaningful results were sent to a cloud service through an Internet of Things framework for further analysis (Čolaković & Hadžialić, 2018).

Along with molecular based monitoring tools, other chemical sensing applications may complement DNA probing (Ishida et al., 2012) and sequencing. An example is provided by *in situ* mass spectrometry, originally developed for targeting xenobiotic compounds in marine water micro-samples, which has been successfully used for identifying species presence based on their physiological by-products (Wollschlager et al., 2016).

### *3.4. Ecosystem indicators*

In the near future, the integration of advanced genomic and chemical approaches for *in situ* detection of organisms (e.g. Cordier et al., 2017) and quantification of their biochemical activity (Goodridge & Valentine, 2016) will greatly enhance the performance of ecological monitoring networks, adding to the detection capacity of optoacoustic imaging and passive acoustic approaches alone.

Stitched imaging products (e.g. mosaicking) can provide valuable information on species distribution and habitat use at more extended scales. *In situ* molecular methods can detect the presence of taxa otherwise undetected by imaging outside a small temporal window or too small for morphological recognition, while acoustics expand the spatial scales of deep-sea biological monitoring, enabling the integration of horizontal (nektobenthic displacement; Aguzzi & Company, 2010) and vertical (i.e. benthopelagic coupling; Griffiths et al., 2017) biomass and energy fluxes. With the use of such combined datasets, a series of biological variables can be measured and ecosystem indicators extracted, as essential elements for the accurate assessment of the health of benthic ecosystems and cover the complete range from benthic (e.g. chemosynthetic) primary production, individual characteristics, population dynamics, species and community dynamics to finally the ecosystem functioning level.

As a result of these ever-growing demands, the need for automation in data collection, analysis and interpretation procedures is paramount. Integration of cabled observatories and associated mobile systems equipped with AI for real-time content extraction from imaging systems, hydrophones and e-DNA samplers, would allow the monitoring of ecosystem indicators and representation of ecosystem functioning over extended spatiotemporal scales (from square meters to kilometers, over days, months, seasons, and decades). To date, no such integrated system exists in the deep sea to verify the concept (Danovaro et al., 2017). At present there are major shortfalls in automation of image and sound processing and producing an efficient, long-term *in situ* e-DNA extraction and sequencing device. However, many of these systems are integrated into the ONC cabled observatory infrastructure in the NE Pacific, with data being collected in real-time at a number of nodes and returned to a central repository (i.e. Ocean 2.0 data bank system). Similarly, real-time, interactive tools such as the Scripps Plankton Camera System (http://spc.ucsd.edu/) facilitate quick access to visual data and a statistical overview. The implementation of these types of data repositories can allow environmental comparisons to be made among neighboring and more distantly arrayed platforms in an attempt to scale local results over a larger networked area (**Figure 4**). This endeavor is providing the guidelines for future development of spatiotemporally integrated monitoring protocols.

Autonomous monitoring of biological variables and derived ecosystem indicators by cabled observatories and their integrated mobile platforms should be implemented following a general and standardized common operational protocol: *i.* all multiparametric readings from optoacoustic imaging, PAM, molecular, biogeochemical, and oceanographic sensors should be acquired synchronously by all cabled and mobile platforms; *ii.* such data acquisition should occur in a high-frequency and time-lapse mode, where the image content should be automatically analyzed by AI algorithms and classified on board of the device, (saving storage PAM and transmission bandwidth space), while preserving the observation time georeferenced stamp; and finally, *iii.* all mobile platforms should constantly survey the same benthic and pelagic areas (subdivided into specific stations) among cabled observatories and their moored vertical projections. Such an automated and spatiotemporally coordinated and standardized protocol for data acquisition will make data treatment, transmission, and storage easier, whilst simultaneously facilitating more straightforward repeatability/reproducibility of observations at the same location and comparison of measurements made with other networks, allowing regional/global level analysis.

The measurement of biological variables needed for the hierarchical computation of ecosystem indicators, should be carried out through a series of sequential automated steps (**Table 2**). *i.* all imaging outputs initially processed for the classification, counting and tracking of fauna and quantification of bacterial mat coverage and activity; *ii.* animals measured (e.g. by stereovision, acoustic scaling or laser scanning) to obtain class-size frequency distribution and sex-ratio (when morphology allows individual discrimination); *iii.* total species counts from all seabed and water column areas summed and standardized for the imaged volume, to obtain an overall abundance (i.e. density) and biomass estimation; *iv.* species counts computed for each station analyzed by mapping procedures (e.g. percentage of presence/occupation per quadrant), to derive information on habitat use as well as displacement routes through different zones (i.e. corridors); *v.* a species richness list and biodiversity obtained at each platform (alpha diversity), between platforms (beta diversity), and the level of the whole network (gamma diversity), to assess habitat heterogeneity influences on species distribution, community composition, and overall ecosystem boundaries; finally, *vi.* density and biomass for each species related to carbon inputs from benthopelagic fluxes in chlorophyll-*a* and turbidity (as proxy for transported organic and inorganic matter), as well as from geochemical fluxes, when relevant (i.e. carrying the reduced chemicals as for example, methane, hydrogen, sulfide, that fuel chemosynthetic microbes), to calculate ecosystem functioning and productivity performances. All automated analysis stages need to be verified by human researchers to ensure accuracy of the algorithm functioning, while the nature of specified ecological interpretation must be cross-checked against published results from conventional methods such as analysis of stomach contents, stable isotopes and fatty acids markers (Choy et al., 2017).

## 4. Perspectives and Outlook

Autonomous flexible networks of cabled observatories and mobile platforms can allow extensive monitoring of marine life at different levels of biological organization and at unprecedented spatial and temporal resolution. Although integrated monitoring actions such as those outlined herein are yet to attain full operational readiness, and therefore proofs of some of the concepts discussed are missing, the technological developments are ongoing. Progress to date already allows researchers to utilize services-oriented ecological monitoring of some isolated deep-sea ecosystems. It is important, however, that future observatories are designed from the ground up for ecosystem monitoring and data integration, rather than being developed on an *ad hoc*, and somewhat haphazard basis, as funding for individual projects becomes available.

Bio-imaging technologies already play a central role in ecosystem exploration and monitoring. Increasing levels of automation in image processing are transforming cameras into true sensors, delivering time series data for a number of biological variables and derived ecosystem indicators. Visual data are being increasingly complemented by *in situ* passive acoustic listening sensors and new e-DNA sequencing technologies for species traceability. All these initially disparate data sources can be combined to form a detailed and high-resolution monitoring approach applicable to the benthic and pelagic components of a deep-sea ecosystem. The output from such a monitoring regime will support decisions of policy makers, allowing them to assess the impacts of increased industrial activities and pressures on deep-sea ecosystems (e.g. oil or gas extraction and mining or trawl fishing), including a better assessment of already evident but poorly quantified climate change impacts at great depths. The obtained data will be of paramount importance for the accurate assessment of the health status of ecosystems, the physical damage to habitats and to efficiently monitor their resilience and the efficacy of restoration actions. The compiling of multiannual time series monitoring data sets (continuously updated in real-time) will allow the identification of shifting environmental baselines and rapidly highlight the onset of any negative environmental impacts which may develop, potentially unpredictably, from human activities in these remote deep-sea ecosystems.

**Funding**

This work was developed within the framework of the Tecnoterra (ICM-CSIC/UPC) and the following project activities: ARIM (Autonomous Robotic sea-floor Infrastructure for benthopelagic Monitoring; MartTERA ERA-Net Cofound), ARCHES (Autonomous Robotic Networks to Help Modern Societies; German Helmholtz Association), RESBIO (TEC2017-87861-R; Ministerio de Ciencia, Innovación y Universidades), RESNEP (CTM2017-82991-C2-1-R; Ministerio de Ciencia, Innovación y Universidades), MERCES (Marine Ecosystem Restoration in Changing European Seas -GA n. 689518) and IDEM (Implementing the MSFD to the Deep Mediterranean Sea, GA No. 11.0661/2017/750680/SUB/ENV.C2).

**Table 1. Biological studies from permanent, autonomous platforms.** Worldwide applications of monitoring of deep-water and deep-sea ecosystems by cabled observatories and crawlers, at different ecologically-ranked levels (i.e. individual, population, community and ecosystem).

| Hierarchical step | Monitored biological variable or ecosystem indicator | Cabled platform | Publications |
|---|---|---|---|
| 1. (Semi-) Automated detection, counting and classification by optoelectronic methods (imaging) | Cold-seep fauna abundance | JAMSTEC Cabled Observatory (1100 m; NW Pacific) | Aguzzi et al. (2009) |
| | | | Aguzzi et al. (2010) |
| | Macrofaunal abundance | LoVe Ocean Observatory (250 m; Norwegian Sea) | Osterloff et al. (2016) |
| | Zooplankton abundance | NEPTUNE Cabled Observatory (ONC; Barkley Canyon; 400-1000 m; NE Pacific) | De Leo et al. (2018) |
| | Benthic fauna abundance, bacterial mat coverage | VENUS Cabled Observatory (ONC; Saanich Inlet; 100 m; NE Pacific) | Aguzzi et al. (2011) |
| | Biological scattering layers classification | | Ross et al. (2013) |
| 2. (Semi-) Automated detection, counting and classification by passive acoustic monitoring (PAM) | Cetacean abundance | ALOHA Cabled Observatory (4700 m; Central Pacific) | Oswald et al. (2011) |
| | | ANTARES Neutrino Telescope (2500 m; Western Mediterranean) | André et al. (2017) |
| | | MARS Cabled Observatory (MBARI; 900 m; NE Pacific) | Ryan et al. (2016) |
| | | NEMO-SN1 (OnDE; EMSO; KM3NeT-It; 2100 m; Central Mediterranean) | Caruso et al. (2015) |
| | | | Caruso et al. (2017) |
| | | NEPTUNE Cabled Observatory (ONC; Endeavour; 2200 m; NE Pacific) | Weirathmueller et al. (2017a) |
| | | NEPTUNE Cabled Observatory (ONC; Cascadia Basin; 2700 m; NE Pacific) | Weirathmueller et al. (2017b) |
| | | NEPTUNE Cabled Observatory (ONC; Barkley Canyon slope; 400-1000 m; NE Pacific) | Kanes et al. (2017) |
| | Zooplankton abundance | | De Leo et al. (2018) |
| 3. Behavioral and life traits (e.g. morphology, rhythms, motility, trophic interactions, territoriality, etc.) | Cetacean diel activity | ALOHA Cabled Observatory (4700 m; Central Pacific) | Oswald et al. (2011) |
| | | ANTARES Neutrino Telescope (2500 m; Western Mediterranean) | André et al. (2017) |

Continued in next page

**Table 1. Biological studies from cabled observatories.** Continued from previous page.

| | | | |
|---|---|---|---|
| 3. Behavioral and life traits (e.g. morphology, rhythms, motility, trophic interactions, territoriality, etc.) | Cold-seep fauna movement, behavior and diel activity | JAMSTEC Cabled Observatory (1100 m; NW Pacific) | Aguzzi et al. (2009) |
| | Cold-seep fauna diel and tidal rhythms | | Aguzzi et al. (2010) |
| | Environmental variability and carcass decomposition | | Aguzzi et al. (2012) |
| | Cetacean diel activity | MARS Cabled Observatory (MBARI; 900 m; NE Pacific) | Ryan et al. (2016) |
| | | NEMO-SN1 (OnDE; EMSO; KM3NeT-It; 2100 m; Central Mediterranean) | Caruso et al. (2017) |
| | Deep-sea inertial bioluminescence rhythms | NEMO Phase-2 tower (KM3NeT-It; 2500 m; Central Mediterranean) | Aguzzi et al. (2017) |
| | Benthic megafauna movement | NEPTUNE Cabled Observatory (ONC; Barkley Canyon shelf-break; 400 m; NE Pacific) | Robert & Juniper (2012) |
| | Benthic fauna movement | NEPTUNE Cabled Observatory (ONC; Barkley Canyon slope; 600-1000 m; NE Pacific) | Chauvet et al. (in Press) |
| | Benthic fauna behavior and diel activity | | Doya et al. (2014) |
| | Benthic fauna diel activity | | Matabos et al. (2014) |
| | Cetacean diel activity | | Kanes et al. (2017) |
| | Hydrothermal vent fauna behavior and diel activity | NEPTUNE Cabled Observatory (ONC; Endeavour; 2200 m; NE Pacific) | Cuvelier et al. (2014) |
| | Hydrothermal vent fauna diel activity | | Cuvelier et al. (2017) |
| | Hydrothermal vent macrofauna rhythms | | Lelièvre et al. (2017) |
| | Cold-seep fauna diel activity | NEPTUNE Cabled Observatory (ONC; Barkley Canyon hydrates; 900 m; NE Pacific) | Chatzievangelou et al. (2016) |
| | Cold-seep fauna behavior | | Doya et al. (2017) |
| | Environmental variability and carcass decomposition | VENUS Cabled Observatory (ONC; Strait of Georgia; 300 m; NE Pacific) | Anderson and Bell (2016) |
| | | VENUS Cabled Observatory (ONC; Saanich Inlet; 100 m; NE Pacific) | Anderson and Bell (2014) |
| | Benthic fauna behavior | | Doya et al. (2016) |
| | Benthic fauna diel activity | | Matabos et al. (2011) |
| | Benthic fauna behavior, diel and tidal rhythms | | Matabos et al. (2015) |
| | Biological scattering layers classification | | Ross et al. (2013) |

Continued in next page

**Table 1. Biological studies from cabled observatories.** Continued from previous page.

| | | | |
|---|---|---|---|
| 4. Population demography, dynamics and distribution (e.g. abundance / biomass, size classes, sex ratio, seasonality, growth and reproduction cycles, spatial distribution, etc.) | Cetacean seasonality | ALOHA Cabled Observatory (4700 m; Central Pacific) | Oswald et al. (2011) |
| | Deep-sea episodic bioluminescence blooms | ANTARES Neutrino Telescope (2500 m; Western Mediterranean) | Tamburini et al. (2013) |
| | Cold-seep bivalve spawning and fecundity | JAMSTEC Cabled Observatory (1200 m; NW Pacific) | Fujikura et al. (2007) |
| | Macrofaunal spatial distribution | LoVe Ocean Observatory (250 m; Norwegian Sea) | Osterloff et al. (2016) |
| | Cetacean seasonality | MARS Cabled Observatory (MBARI; 900 m; NE Pacific) | Ryan et al. (2016) |
| | | NEMO-SN1 (OnDE; EMSO; KM3NeT-It; 2100 m; Central Mediterranean) | Caruso et al. (2017) |
| | | | Sciacca et al. (2015) |
| | Cetacean size estimation | | Caruso et al. (2015) |
| | Benthic megafauna abundance and size | NEPTUNE Cabled Observatory (ONC; Barkley Canyon shelf; 400 m; NE Pacific) | Robert & Juniper (2012) |
| | | DELOS observatories (1400 m, SW Atlantic off Angola) | Vardaro et al. (2013) |
| | Benthic fauna seasonal and inter-annual trends | NEPTUNE Cabled Observatory (ONC; Barkley Canyon slope; 400-1000 m; NE Pacific) | Chauvet et al. in press |
| | Zooplankton seasonal and interannual trends | | De Leo et al. (2018) |
| | Benthic fauna size classes and migrations | | Doya et al. (2014) |
| | Benthic fauna seasonality | | Juniper et al. (2013) |
| | | DELOS observatories (1400 m, SW Atlantic off Angola) | Vardaro et al. (2013) |
| | Cold-seep fauna spatial distribution | NEPTUNE Cabled Observatory (ONC; Barkley Canyon hydrates; 900 m; NE Pacific) | Chatzievangelou et al. (2017) |
| | Cold-seep fauna seasonality and reproductive cycles | | Doya et al. (2017) |
| | Cold-seep fauna seasonality | | Thomsen et al. (2017) |
| | Hydrothermal vent fauna microhabitat use | NEPTUNE Cabled Observatory (ONC; Endeavour; 2200 m; NE Pacific) | Cuvelier et al. (2017) |
| | | | Cuvelier et al. (2014) |
| | Cetacean seasonality, inter-annual trends and spatial distribution | | Weirathmueller et al. (2017a) |
| | Benthic fauna size classes, seasonality and microhabitat use | VENUS Cabled Observatory (ONC; Saanich Inlet; 100 m; NE Pacific) | Doya et al. (2016) |
| | Benthic fauna size classes and seasonality | | Matabos et al. (2012) |
| | Biological scattering layers classification | | Ross et al. (2013) |

Continued in next page

**Table 1. Biological studies from cabled observatories.** Continued from previous page.

| | | | |
|---|---|---|---|
| 5. Biodiversity (e.g. composition, richness, alpha / beta / gamma diversity, etc.) | Cold-seep fauna diversity | NEPTUNE Cabled Observatory (ONC; Barkley Canyon hydrates; 900 m; NE Pacific) | Chatzievangelou et al. (2017) |
| | Benthic fauna diversity | NEPTUNE Cabled Observatory (ONC; Barkley Canyon slope; 900-1000 m; NE Pacific) | Chauvet et al. (in Press) |
| | | | Juniper et al. (2013) |
| | Benthic community composition | | Matabos et al. (2014) |
| | | DELOS observatories (1400 m, SW Atlantic off Angola) | Vardaro et al. (2013) |
| | Hydrothermal vent fauna community composition | NEPTUNE Cabled Observatory (ONC; Endeavour; 2200 m; NE Pacific) | Cuvelier et al. (2017) |
| | Benthic community composition | VENUS Cabled Observatory (ONC; Saanich Inlet; 100 m; NE Pacific) | Matabos et al. (2012) |
| | Benthic community dynamics | | Matabos et al. (2015) |
| 6. Ecosystem functioning (food-web structure, carbon flux, bioturbation /remineralization) | Seabed bioturbation by benthic megafauna | NEPTUNE Cabled Observatory (ONC; Barkley Canyon shelf-break; 400 m; NE Pacific) | Robert & Juniper (2012) |
| | Zooplankton carbon fluxes | NEPTUNE Cabled Observatory (ONC; Barkley Canyon slope; 400-1000 m; NE Pacific) | De Leo et al. (2018) |
| | Seasonal carbon fluxes | NEPTUNE Cabled Observatory (ONC; Barkley Canyon hydrates; 900 m; NE Pacific) | Thomsen et al. (2017) |

**Table 2. Indicators extraction roadmap.** Consecutive automatable steps for the hierarchical computing of ecosystem indicators from input biological variables, obtained by bio-imaging and other sensing technologies, installed on spatially distributed autonomous networks of cabled observatories and their connected mobile benthic and pelagic platforms.

| | | |
|---|---|---|
| Biological variables | ***Platform level:*** **Input data from each field of view** | |
| | ***Animal tracking and classification*** | •Automated processing of video and acoustic imaging: faunal identification and counting<br>•Automated processing of passive acoustic readings: detection and classification of organisms<br>•Complementary e-DNA molecular sequencing: species presence traceability |
| | ***Demographic and functional descriptors*** | •Class-size frequencies distributions (e.g. by laser scaling or sterovision)<br>• Morphology, colour<br>•Behavioural life traits (e.g. territoriality, mode of displacement, rhythmic activity, predatory modality) |
| | ***Network level:*** **Data standardization over total imaged surface/volume** | |
| | ***Biometric Population dynamics*** | •Abundance<br>•Biomass<br>•Body size and growth cycles (e.g. moulting, recruiting)<br>•Reproductive cycles (e.g. spawning site and timing) |
| | ***Habitat and species distributions*** | •Seascapes changes (i.e. photomosaics and 3D laser scans)<br>•Percentage of presence/occupation per quadrant (e.g. heat-mapping procedures) |
| Ecosystem indicators | ***Biodiversity*** | •Species composition<br>•Richness of taxa<br>•Alpha (within observational platform), Beta (between platforms), and Gamma (at the level of the entire network) |
| | ***Functioning and productivity*** | •Species associations as proxy of inter-specific relationships (i.e. food web structure)<br>•Abundance, biomass, and faunal behavioural activity as proxy for carbonflux<br>•Bioturbation as proxy for remineralization |

**Figure legends**

**Figure 1.** The fixed (i.e. cabled) and mobile docked platforms constituting a spatial network for the integrated benthic and pelagic ecosystem monitoring. (A) Video-cabled multiparametric observatory platform, acting as a docking station for a pelagic Remotely Operated Vehicle (ROV) and a tethered mobile benthic crawler (Courtesy of Dr. O. Godø & Dr. T. Torkelsen); (B) rover (MANSIO-VIATOR) similar to crawlers but not tethered, docked to a vessel-assisted repositioning station; (C) Architecture of ANTARES (the Astronomy with a Neutrino Telescope and Abyss environmental RESearch detector) with a line of Photo-Multiplier Tubes (PMTs) and a tethered crawler.

**Figure 2.** Illustration of a variety of cabled observatories providing the sea bed infrastructure to control and coordinate mobile benthic and pelagic platforms such as docked crawlers, rovers, and AUVs. Platform monitoring is assisted by vessels and satellite-based technologies. Neut. Telescope – is an array of vertical moored lines of Photo-Multiplier Tubes (PMTs) deployed in the deep sea. Seabed infrastructures providing power and data transfer may be aided by connection with industrial or telecommunication cables, as reliable low-cost means for network deployment into vast abyssal areas (Danovaro et al., 2017).

**Figure 3.** Different video and acoustic imaging data outputs obtained by fixed-point and crawler platforms connected through the Ocean Networks Canada (ONC). (A, B) Commercially exploited sablefish (*Anoplopoma fimbria*) imaged from cabled observatory HD video (A) and from an imaging rotary sonar (B), at 970 m depth in Barkley Canyon (NE Pacific); (C) A spiny dogfish (*Squalus achantias*) imaged with the ARIS acoustic camera at 120 m depth from a cabled observatory in the Strait of Georgia. The color scale bar indicates raw backscatter reflectivity amplitude (in decibels, dB) (courtesy of X. Mouy); (D, E) Photomosaic panoramas obtained by the crawler in Barkley Canyon (870 m; NE Pacific, Canada), depicting egg towers of *Neptunea* sp. snails (D), and a range of benthic species occupying a methane seep habitat patch; (F,G) 3D photomosaics of a methane hydrate mound at the same location, depicting mound area/volume changes over time due to uplift/growth in hydrates (areas in yellow) and slumping (areas in red; courtesy of Dr. T. Kwasnitschka).

**Figure 4.** Schematic representation of current seafloor monitoring infrastructure ONC in Barkley Canyon, where a power node distributes energy and data transmission capability to serve fixed multiparametric imaging-platforms and a crawler. The mobile platforms communication and coordinated function makes this area the first cooperative network (shown in the bubble) for the deep-sea ecological monitoring. As an example of the power of ongoing multiparametric monitoring, time series from several environmental sensors for the crawler are presented over consecutive years (data plotted at 1-h frequency). When gaps in data acquisition occur, data can be supplied by nearby cabled platform (as interpolated to cover maintenance periods) Shaded green areas indicates moments at which the environmental monitoring by the crawler has been accompanied by image collection, the processing of which is still manual, while automated scripts for animal tracking and species classification are under development.

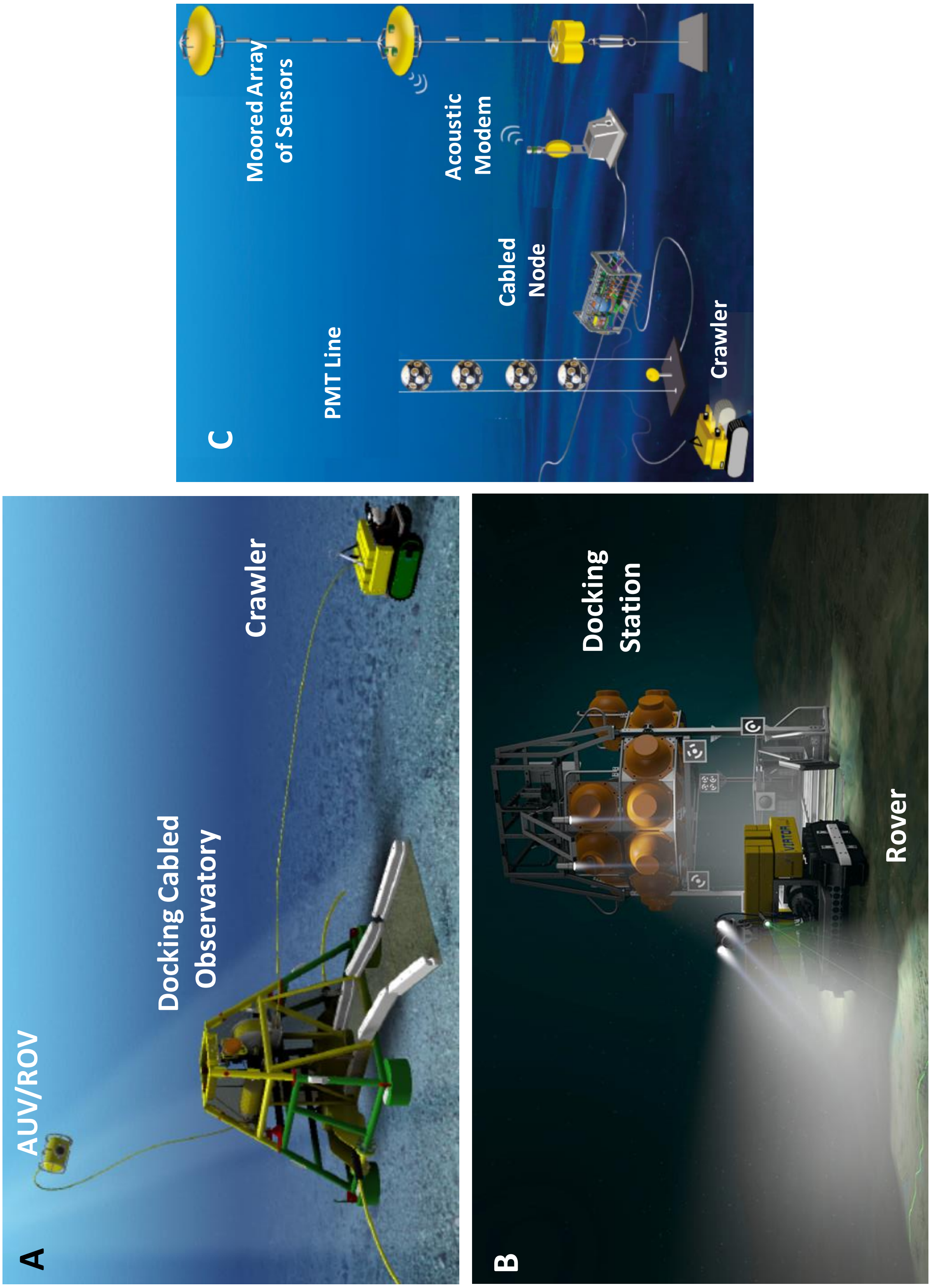

**Aguzzi *et al.* Figure 1**

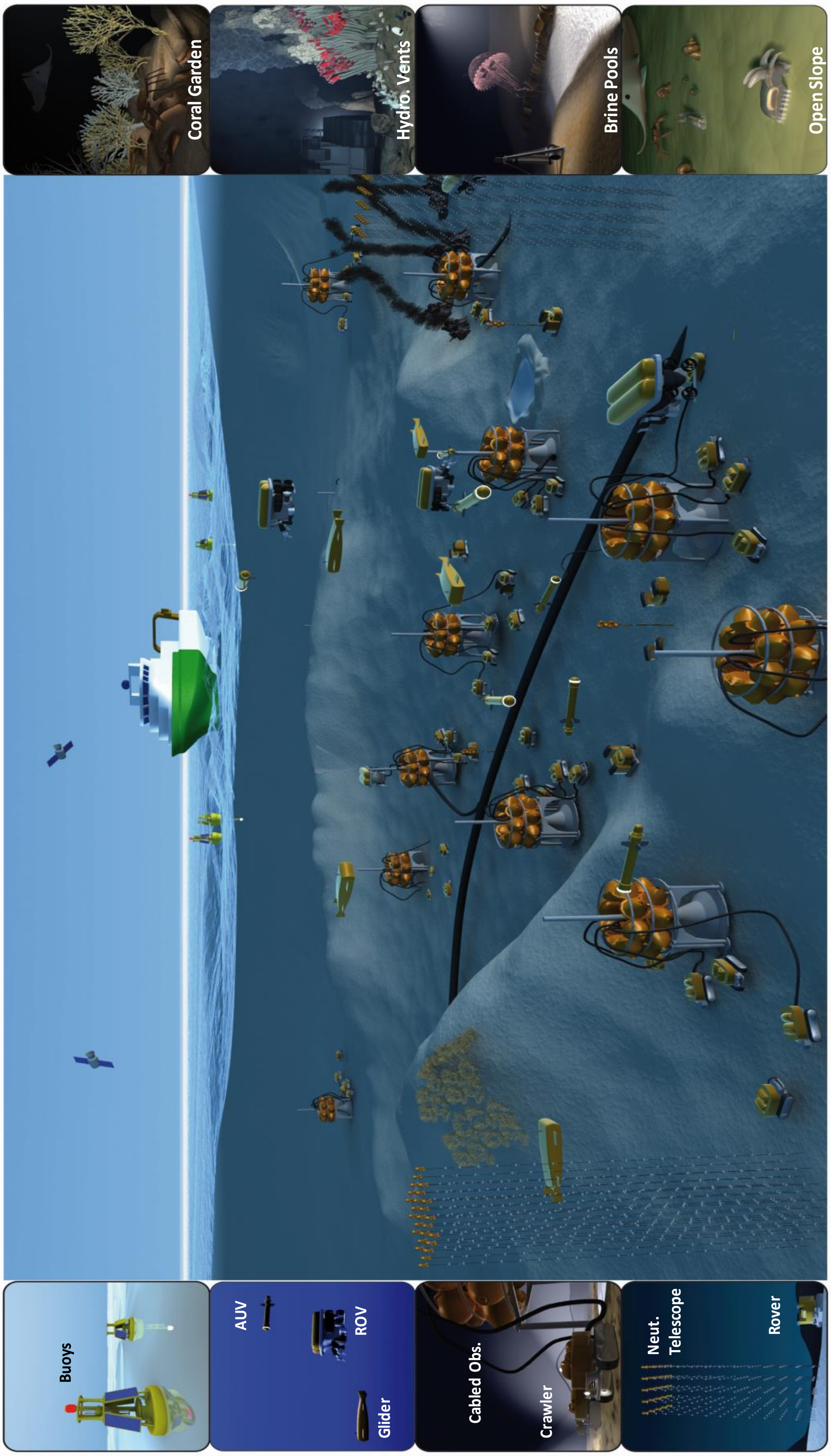

**Aguzzi *et al.* Figure 2**

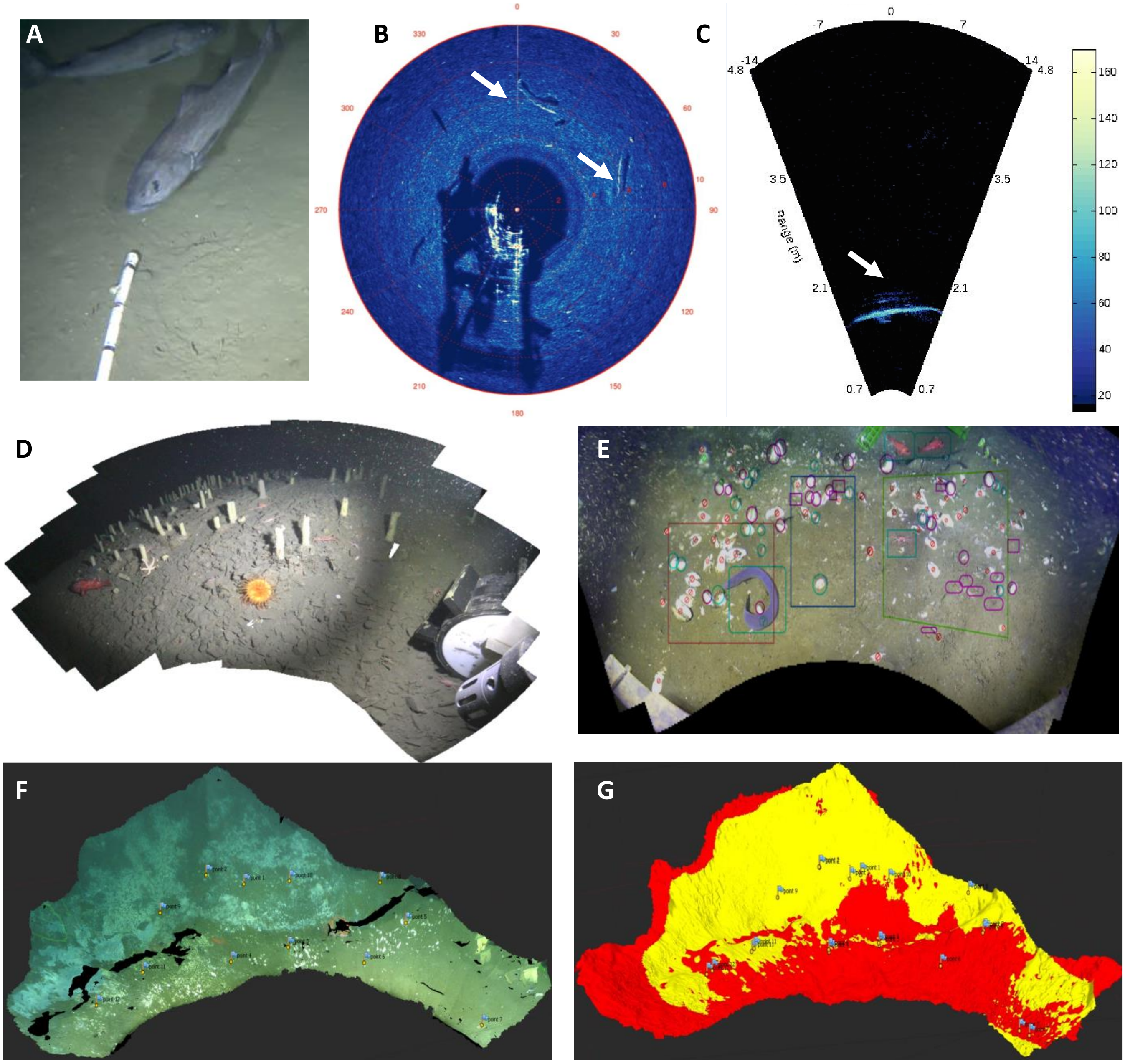

Aguzzi *et al.* Figure 3

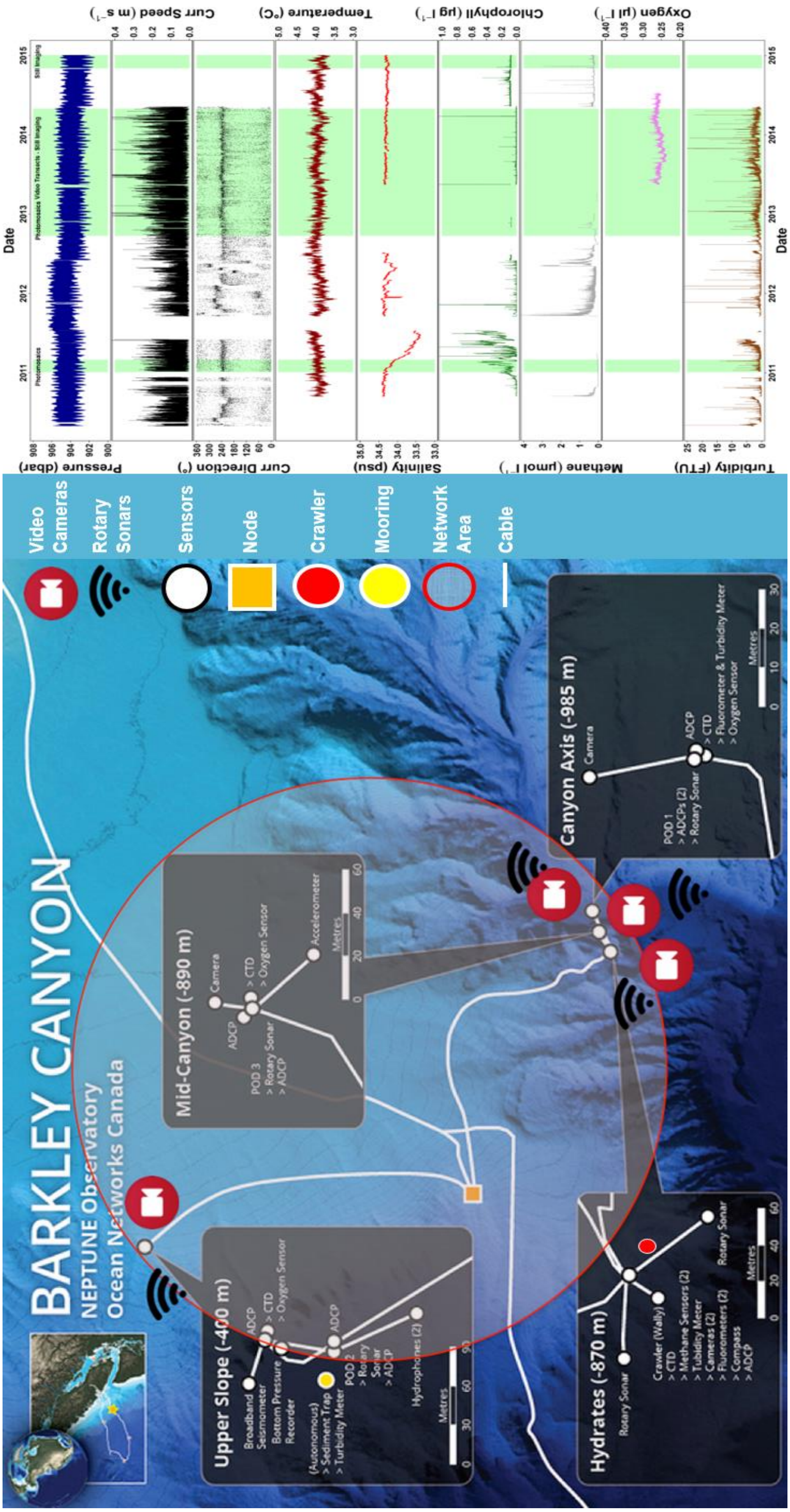

**Aguzzi *et al.* Figure 4**